\documentclass[12pt]{article}

\def\rdots{\mathinner{\mkern1mu\raise1pt\vbox{\kern1pt\hbox{.}}\mkern2mu
   \raise4pt\hbox{.}\mkern2mu\raise7pt\hbox{.}\mkern1mu}}
\newcommand{\Z}{{\rm Z\kern-.35em Z}}
\newcommand{\bP}{{\rm I\kern-.15em P}}
\newcommand{\Q}{\kern.3em\rule{.07em}{.65em}\kern-.3em{\rm Q}}
\newcommand{\R}{{\rm I\kern-.15em R}}
\newcommand{\h}{{\rm I\kern-.15em H}}
\newcommand{\C}{\kern.3em\rule{.07em}{.65em}\kern-.3em{\rm C}}
\newcommand{\T}{{\rm T\kern-.35em T}}

\newcommand{\be}{\begin{equation}}
\newcommand{\ee}{\end{equation}}

\newcommand{\ra}{\rightarrow}

\newcommand{\nn}{\nonumber}

\begin{document}
  
\openup 1.5\jot
\centerline{Linear Inflation in Curvature-Quadratic Gravity}

\vspace{1in}
\centerline{Paul Federbush}
\centerline{Department of Mathematics}
\centerline{University of Michigan}
\centerline{Ann Arbor, MI 48109-1109}
\centerline{(pfed@math.lsa.umich.edu)}

\vspace{1in}

\centerline{\underline{Abstract}}

We continue our study of gravity described by the action density $(-g)^{1/2} (R_{ik} R^{ik} + bR^2)$; and look for cosmological solutions of gravity coupled to dust, for the closed isotropic model.  There is a solution that at $t \ra 0$ has for the radius
, $a(t) = t/\sqrt{3}$; in the absence of dust this solution holds for all time.

\vfill\eject

In a previous paper, [1], we have proposed the curvature-quadratic action
\be	\frac 1{2c} \int d^4x(-g)^{1/2} [R^{ik} R_{ik} + bR^2]	\ee
as the basis for quantum gravity.  If $b = - \frac 1 3$ then this action is of the conformal Weyl theory.  We have already studied classical solutions arising from (1) of the Schwarzschild form [2], and in this paper turn to cosmological solutions of the 
closed isotropic form.

We write the metric in usual form:
\be
ds^2 = -dt^2 + a^2(t) \Big[ d \chi^2 + \sin^2 \chi (d\theta^2 + \sin^2 \theta d\phi^2) \Big].
\ee
For the Einstein action
\be	\frac 1{16\pi k} \int d^4 x(-g)^{1/2} \ R \ .	\ee
The equation for $a(t)$ is obtained as:
\be	3 \left( \frac{ \dot{a}^2} {a^2} +  \frac 1{a^2}  \right)  = (8\pi k)\rho \ .	\ee 
(See [3], for example).  For the action of (1) equation (4) is replaced by
\begin{eqnarray}
Q\Big[ &-& 18 + 18 a^2 (a'')^2 + 54(a')^4 + 36(a')^2  \\
&-& 35(a')^2 a a'' - 36 a' a^2 a''' \Big] = -ca^4 \rho \nn
\end{eqnarray}
We have written
\be		b= - \frac 1 3 + Q 	\ee
and assume $Q \not= 0$.  There is the conservation law
\begin{eqnarray}
\rho a^3 &=& \ \ {\rm const.} \\
&\equiv& M \ . \nn 
\end{eqnarray}
If $\rho = M = 0$, we see that a solution of (5) is
\be	a(t) = \frac 1 {\sqrt{3}} \ t \ .		\ee
If $M\not= 0$, we write a solution in terms of the series
\be
a(t) = \frac 1 {\sqrt{3}} \ t \  + c_2t^2 + c_3t^3 + \dots \ .	\ee
Substituting (9) into (5) we find the first few terms:
\begin{eqnarray}
c_2 &=& \frac{-1}{264} \cdot \left( \frac{cM}{Q} \right) \\
c_3 &\cong& -1.87 \cdot 10^{-5} \cdot \left( \frac{cM}{Q} \right)^2 \ .
\end{eqnarray}
We do not discuss how incredibly small the effective $c$ would have to be for the current inflation to be physically relevant.

If $Q=0$ one is in the conformal situation, which is rather different and not considered here at all.  Since the work of Starobinsky in [4] there has been much work on the possibility of higher order terms in gravity leading to inflation.  But we are not 
aware that the results of the present note have been pointed out.

\bigskip
\bigskip

\centerline{\underline{References}}

\begin{itemize}
\item[[1]] P. Federbush, ``Speculative Approach to Quantum Gravity", Symposium in Honor of Eyvind H. Wichmann, University of California, Berkley, June 1999.
\item[[2]] P. Federbush, ``On Schwarzschild-Like Solutions in Curvature-Quadratic Gravity", gr-gc/0002037.
\item[[3]] L. Landau and E. Lifshitz, \underline{The Classical Theory of Fields}, Addison-Wesley Press (1951)
\item[[4]]  A.A. Starobinsky, ``A New Type of Isotropic Cosmological Models without Singularity", {\it Phys. Lett.}, B91, 99 (1980).
\end{itemize}

\end{document}